\documentclass[
preprint,
 amsmath,amssymb,
 aps,
showkeys,
showpacs
]{revtex4-2}

\usepackage{comment}
\usepackage{graphicx}
\usepackage{dcolumn}
\usepackage{bm}

\newcommand{\be}{\begin{eqnarray}}
\newcommand{\ee}{\end{eqnarray}}

\begin{document}

\preprint{MUPB/Conference section: Dark Matter theory}

\title{New Options for SUSY-kind Dark Matter }
\thanks{Supported by the RSF Grant 20-42-09010}%

\author{Elena Arbuzova}
 \email{arbuzova@uni-dubna.ru}
\affiliation{%
 Department of Higher Mathematics, Dubna State University,\\ Universitetskaya st. 19, Dubna, 141983 Russia\\
 Department of Physics, Novosibirsk State University, \\Pirogova st. 2, Novosibirsk, 630090 Russia 
}%

\date{25 August 2021}
             
\begin{abstract}
In the conventional cosmology masses of the stable supersymmetric
relics, candidates for the dark matter (DM) particles, should be typically below 1 TeV. This is in conflict
with the LHC bounds on the low energy SUSY. However, in $R^2$-gravity the masses
of the stable particles with the interaction strength typical for SUSY could be much higher
depending upon the dominant decay mode of the scalaron. We discuss the bounds on the masses of
DM particles for the following dominant decay modes: to minimally coupled massless scalars, to
massive fermions, and to gauge bosons.
\end{abstract}

\keywords{dark matter, $R^2$-cosmology, scalaron decay}
\pacs{04.50.Kd, 95.35.+d}
\maketitle

One of the most intriguing problem of modern cosmology is the mystery of dark matter. DM is invisible form of matter disclosing itself through its gravitational action. 
An accepted property of DM particles is that they are electrically neutral, since do not scatter light. Other their properties are practically unknown and these circumstances open possibility for particles of many different types to be dark matter candidates. 

A natural, and formerly very popular, candidate to be the DM constituent is the lightest supersymmetric particle (LSP). Low energy supersymmetric model predicts the existence of stable LSP with a mass of several hundred GeV. However the absence of manifestation of supersymmetric partners at LHC considerably restricts the parameter space open for SUSY. 

The cosmological energy density of LSPs is proportional to their mass squared:
\be 
\rho_{LSP} \sim \rho_{DM}^{(obs)} (M_{LSP}/ 1\,{TeV})^2.  
\label{rho-LSP}
\ee
One can see from this expression that 
for LSPs with the mass $M_{LSP} \sim 1$ TeV
 their energy density is of the order of the observed DM energy density, $\rho_{DM}^{(obs)} \approx 1\ keV/cm^3$, but for larger masses LSPs would overclose  the universe. Thereby LSPs are practically excluded as DM particles in the conventional cosmology. However,  the energy density of LSPs may be much lower in 
 $(R+R^2)$-gravity with the action
\be
S_{tot} = -\frac{M_{Pl}^2}{16\pi} \int d^4 x \sqrt{-g} \left(R-\frac{R^2}{6M^2_{R}}\right)+S_m\,, 
\label{S-R2-tot-1}
\ee 
where $M_{Pl}=1. 2 2\cdot 10^{19}$ GeV is the Planck mass. Curvature ${ R(t)}$ can be considered as an effective scalar field (scalaron) with the mass $M_R$ and with the decay width $ \Gamma $. In model \eqref{S-R2-tot-1} the density of heavy relics  could be
significantly reduced by radiation from the scalaron decay and it reopens for LSPs the chance  to form dark matter, if   $M_{LSP} \geq 1000$TeV \cite{Arbuzova:2018apk}.  

Added to the usual Einstein-Hilbert action $R^2$-term leads to exponential cosmological expansion  (Starobinsky inflation) \cite{Starobinsky:1980te}, creates considerable deviation from the Friedmann cosmology in the post-inflationary epoch \cite{Arbuzova:2018ydn} and changes significantly kinetics of massive species and the density of dark matter particles \cite{Arbuzova:2021etq}. 

Cosmological energy density of the produced particles depends upon the dominant decay channel of the scalaron. 
If scalaron decays into 2 massless scalars minimally coupled to gravity, the decay width and the energy density are respectively: 
\be 
\Gamma_s = \frac{M_R^3}{24M_{Pl}^2}, \ \ \ \   \rho_{s} = \frac{M_R^3}{240 \pi t}.
\label{Gamma-s}
\ee
For scalaron decay into a pair of fermions  with mass ${m_f}$ we have: 
\be 
\Gamma_f = \frac{ M_R m_f^2 }{6 M_{Pl}^2 }, \ \ \ \ \rho_f = \frac{ M_R m_f^2}{240 \pi t}. 
\label{Gamma-f}
\ee
If the scalaron decay is induced by the conformal anomaly, the decay width of the scalaron and the energy density of the produced gauge bosons will be correspondingly \cite{Arbuzova:2020etv}:
 \be 
\Gamma_{an} = \frac{\beta_1^2 \alpha^2 N}{96\pi^2}\,\frac{M_R^3}{M_{Pl}^2} , \ \ \ \ 
\rho_{an} = \frac{\beta^2_1 \alpha^2 N}{4 \pi^2} \,\frac{M_R^3}{120 \pi t},
\label{Gamma-an}
\ee
{where
$\beta_1$ is the first coefficient of the beta-function, $N$ is the rank of the gauge group, 
$ \alpha$ is the gauge coupling constant (at high energies it depends upon the model).}

Let us consider the evolution of massive species $X$ in plasma with temperature $T$. 
The number density $n_X$~of particles having mass $M_X$ is governed by the Zeldovich equation~\cite{Zeldovich:1965gev}:
\be 
\dot n_X + 3H n_X = -\langle \sigma_{ann} v \rangle \left( n_X^2 - n^2_{eq} \right), \   
n_{eq} = g_s \left(\frac{M_X T}{2\pi}\right)^{3/2} e^{-M_X/T}, 
\label{dot-n-X}
\ee
where $ \langle \sigma_{ann} v \rangle$ is the thermally averaged annihilation cross-section of X-particles,
$n_{eq}$ is their equilibrium number density, $g_s$ is the number of spin states, $H$ is the Hubble parameter.

For annihilation of the non-relativistic particles:
\be
\langle \sigma_{ann} v \rangle&=& \sigma_{ann} v = \frac{\pi \alpha^2 \beta_{ann}}{2M_X^2} \ \ \text{(S-wave)},\\ \nonumber
\langle \sigma_{ann} v \rangle &=&  \frac{3\pi \alpha^2 \beta_{ann}}{2M_X^2} \,\frac{T}{M_X} \ \ \text{(P-wave, Majorana fermions)},
\label{sigma-ann}
\ee 
where $\alpha$ is a coupling constant, in SUSY theories ${\alpha \sim 0.01}$, and ${\beta_{ann}}$ is a numerical parameter proportional to 
the number of annihilation channels, ${\beta_{ann}}\sim 10$. 

We assume that the direct $X$-particle production by curvature ${R(t)}$ 
is suppressed in comparison with the inverse annihilation of light particles into $X\bar X$-pair. 

We have solved the Zeldovich equation \eqref{dot-n-X} for different decay channels of the scalaron  (see Eqs.~\eqref{Gamma-s}--\eqref{Gamma-an})
and found the frozen energy density of the produced $X$-particles, $\rho_X$.  Comparing it with the observed energy density of dark matter, ${ \rho_{DM}^{(obs)} \approx 1}$ keV/cm$^3$, we have shown that the range of the allowed masses of $X$-particles to form the cosmological dark matter significantly depends upon the dominant decay mode of the scalaron. The results are presented in Table~\ref{tab:table1}. 
\begin{table}[b]
\caption{\label{tab:table1}%
The range of the allowed masses of $X$-particles to form  cosmological dark matter 
}
\begin{ruledtabular}
\begin{tabular}{lcdr}
\multicolumn{1}{c}{\textrm{Dominant decay channel of the scalaron}}&
\textrm{Allowed $M_X$ to form DM}\\
\colrule
Minimally coupled massless scalars mode & $M_X \gtrsim M_R \approx 3\cdot 10^{13}$ GeV  \\
Massive fermions mode & $M_X \sim 10^6$ GeV \\
Gauge bosons mode & $M_X \sim 5 \cdot 10^{12}$ GeV \\
\end{tabular}
\end{ruledtabular}
\end{table}

According to our calculations, the mass of DM particles with the interaction strength typical for supersymmetric ones can
be in the range from $10^6$ to $10^{13} $ GeV. It is tempting to find how they could be observed except for their gravitational effects on galactic and cosmological scales. There are some possibilities to make such $X$-particles visible:

1. Annihilation effects in clusters of dark matter in galaxies and galactic halos, where, according to Ref.~\cite{clusters}, the density of DM is much higher than DM cosmological density.

2. The decay of superheavy DM particles, which should have a lifetime long enough to manifest themselves as stable dark matter, but at the same time lead to the possibly observable contribution to the  UHECR spectrum~\cite{uhecr-DM-decay}. 

3. It worth noting that instability of superheavy DM particles can arise due to 
Zeldovich mechanism through virtual black holes formation \cite{zeld-BH-eng}. 

In conclusion we want to emphasize, that the existence of stable particles
with interaction strength typical for SUSY and heavier than several TeV is in tension with conventional Friedmann cosmology. Starobinsky inflationary model opens a way to save life of such $X$-particles, because in this model the density of heavy relics  could be
significantly reduced. If the epoch of the domination of the curvature oscillations (scalaron domination) lasted after freezing of massive species, their
density with respect to the plasma entropy could be noticeably diluted by  radiation from the scalaron decay. The range of allowed masses of $X$-particles to form dark matter depends upon the dominant decay mode of the scalaron.

\begin{acknowledgments}
The work was supported by the RSF grant 20-42-09010.
\end{acknowledgments}


\end{document}